\documentclass[twocolumn,pre,aps,nofootinbib,showpacs]{revtex4}

\usepackage[bookmarks=false]{hyperref}
\usepackage{graphicx}
\graphicspath{{./fig/}}
\newcommand{\BoldVec}[1]{\mathchoice%
  {\mbox{\boldmath $\displaystyle     #1$}}%
  {\mbox{\boldmath $\textstyle        #1$}}%
  {\mbox{\boldmath $\scriptstyle      #1$}}%
  {\mbox{\boldmath $\scriptscriptstyle#1$}}%
}
\newcommand{\EQ}{\begin{equation}}
\newcommand{\EN}{\end{equation}}
\newcommand{\EQA}{\begin{eqnarray}}
\newcommand{\ENA}{\end{eqnarray}}
\newcommand{\eq}[1]{(\ref{#1})}

\newcommand{\Eq}[1]{Eq.~(\ref{#1})}
\newcommand{\Eqs}[2]{Eqs~(\ref{#1}) and~(\ref{#2})}

\newcommand{\Sec}[1]{Sec. \ref{#1}}

\newcommand{\Fig}[1]{Fig.~\ref{#1}}

\newcommand{\Tab}[1]{Table~\ref{#1}}

\newcommand{\bra}[1]{\langle #1\rangle}

\newcommand{\vvv}{\hat{\mbox{\boldmath $v$}} {}}

\newcommand{\xx}{\BoldVec{x}{}}

\newcommand{\vv}{\BoldVec{v} {}}

\newcommand{\kk}{\BoldVec{k} {}}

\newcommand{\nab}{\BoldVec{\nabla} {}}

\newcommand{\dd}{{\rm d} {}}
\newcommand{\const}{{\rm const}  {}}

\def\Rey{\mbox{\rm Re}}

\def\half{{\textstyle{1\over2}}}

\newcommand{\yjfm}[3]{, J. Fluid Mech. {\bf #2}, #3 (#1).}

\newcommand{\ypr}[3]{, Phys.\ Rev.\ {\bf #2}, #3 (#1).}
\newcommand{\yprl}[3]{, Phys.\ Rev.\ Lett.\ {\bf #2}, #3 (#1).}
\newcommand{\yphl}[3]{, Phys.\ Lett.\ {\bf #2}, #3 (#1).}

\newcommand{\yjcp}[3]{, J. Comp. Phys. {\bf #2}, #3 (#1).}

\newcommand{\ypf}[3]{, Phys. Fluids {\bf #2}, #3 (#1).}

\newcommand{\ybook}[3]{ {\em #2}. #3 (#1).}

\begin{document}
\preprint{NORDITA 2003-98 AP}

\title{Self-similar scaling in decaying numerical turbulence}

\author{Tarek A.\ Yousef}
  \email{Tarek.Yousef@mtf.ntnu.no}
  \affiliation{Department of Energy and Process Engineering,
  Norwegian University of Science and Technology,
  Kolbj{\o }rn Hejes vei 2B, N-7491 Trondheim, Norway}
\author{Nils Erland L.\ Haugen}
  \email{Nils.Haugen@phys.ntnu.no}
  \affiliation{Department of Physics, The Norwegian University of Science
  and Technology, H{\o}yskoleringen 5, N-7491 Trondheim, Norway}
\author{Axel Brandenburg}
  \email{Brandenb@nordita.dk}
  \affiliation{NORDITA, Blegdamsvej 17, DK-2100 Copenhagen \O, Denmark}

\date{\today,~ $ $Revision: 1.86 $ $}

\begin{abstract}
Decaying turbulence is studied numerically using as initial condition
a random flow whose shell-integrated energy spectrum increases with
wave number $k$ like $k^q$.
Alternatively, initial conditions are generated from a driven turbulence
simulation by simply stopping the driving.
It is known that the dependence of the decaying energy spectrum on
wave number, time, and viscosity can be collapsed onto a unique scaling
function that depends only on two parameters.
This is confirmed using three-dimensional simulations and the dependence
of the scaling function on its two arguments is determined.
\end{abstract}

\maketitle

\section{Introduction}\label{sec:intro}

According to the classical Kolmogorov theory of 1941 \cite{Kolmogorov41},
hydrodynamical turbulence is an example of a system that is self-similar,
i.e., the velocity pattern is supposed to look
similar when viewed at different degrees
of magnification \cite{Mandelbrot74}.
Of course, real turbulence is not precisely self-similar because
of intermittency effects that are responsible for anomalous scaling,
but for the present purpose such corrections can be regarded as small. 

Two different self-similarity behaviors have been discussed in the
literature: inertial range self-similarity and infrared asymptotic
self-similarity that we shall be concerned with here.
The most famous one is probably the inertial range self-similarity.
Kolmogorov \cite{Kolmogorov41} showed that the velocity difference
between two points, $\delta v$, increases with scale $\ell$ such that
$\langle\delta v \rangle\propto\ell^h$, where $h=1/3$; see also Ref.~\cite{Frisch}.
In other words, when looking at the velocity at a magnified scale,
$\xx\to\alpha\xx$, where $\alpha$ is the magnification factor, then
velocities will only be similar if they are rescaled by a factor $\alpha^h$,
i.e.\ $\vv\to\alpha^h\vv$.

However, at sufficiently small scales, viscous dissipation always 
destroys the self-similarity.
This implies that there will be a modification to an otherwise perfect
power law behavior of the shell integrated energy spectrum, $E(k)$.
This modification can be described by a universal scaling function
$\psi(k,\nu)$, which depends on the kinematic viscosity $\nu$.
Thus, one can write
\EQ
E(k,\nu)=k^q\psi(k,\nu)\quad\text{(forced turbulence)},
\label{ForcedTurbulence}
\EN
where $q=-(1+2h)$ follows from the normalization
$\int E(k,\nu)\,\dd k=\half\bra{\vv^2}$.
For sufficiently large Reynolds numbers,
the energy spectrum has an inertial range with $h=1/3$, i.e., $q=-5/3$.
This spectrum cuts off at the wave number
$k_{\rm d}=(\epsilon/\nu^3)^{1/4}$,
where $\epsilon$ is the rate of energy input.
This dependence on $\nu$ can be used to simplify the scaling function
to a function that has only one argument, i.e.,
\EQ
\psi(k,\nu)=f(k/k_{\rm d})=f(k\nu^{3/4}\epsilon^{-1/4}).
\EN
Here, $f$ is a universal function that depends, in addition to $k$,
only on the outer scale determined by the geometry of the system.

We now discuss the infrared asymptotic self-similarity, i.e.,
in the following the scaling exponents $h$ and $q$ apply no longer
to the inertial range, but to the subinertial (infrared) range.
In the case of decaying turbulence, the scaling function also
depends on time, i.e., $\psi=\psi(k,t,\nu)$.
Furthermore, $\psi$ is not {\it a priori}
universal in the sense that its form may
depend on the initial spectrum.
The spectrum also becomes time dependent,
\EQ
E(k,t,\nu)=k^q\psi(k,t,\nu)\quad\text{(decaying turbulence)},
\label{DecayingTurbulence}
\EN
where $q$ depends on the initial condition.
If the initial condition is restricted to be turbulent so that, prior to
turning off the forcing, the energy spectrum satisfies \Eq{ForcedTurbulence},
$q$  is expected to be somewhere between $1$ and $4$; see
Refs.~\cite{Hinze,Lesieur}.

In a recent paper \cite{DJO}, Ditlevsen, Jensen, and Olesen found that
the scaling function $\psi(k,t,\nu)$
reduces to a two-parametric dependence,
\EQ
\psi(k,t,\nu)=g(kt^a,\nu t^b),
\label{dependence}
\EN
where $g=g(x,y)$ is a new scaling function that has only two arguments,
and $a$ and $b$ are exponents that depend only on the slope of the
infrared part of the initial spectrum.

Using data from decaying wind tunnel turbulence \cite{CBC} it was possible to
show \cite{DJO} that the energy spectra for different times can be collapsed
onto a single graph by plotting $k^{-q}E(k,t,\nu)$ versus $kt^a$.  The
dependence on the viscosity $\nu$, and hence on the second argument $y$ of the
scaling function $g(x,y)$, has been discarded.  This may be appropriate in the
large Reynolds number limit.

The purpose of the present paper is to determine, using numerical
simulations, the dependence of $g(x,y)$ on both $x$ and $y$.
In a first step we determine the dependence on $x$ by keeping
$y$ constant.
This is accomplished by letting $\nu$ vary in such a way that
$\nu t^b\equiv y=\const$. This can obviously not easily be done in
wind tunnel turbulence (although $\nu$
could in principle be changed by varying the temperature).
In a simulation, changing $\nu$ is of course quite straightforward.
The dependence on $y$ is determined by integrating over $x$
and considering the decay law of kinetic energy.

\section{Scaling in decaying turbulence}\label{sec:decaying}

We first recapitulate the derivation presented in Ref.~\cite{DJO}.
The unforced incompressible  Navier-Stokes equation
\EQ\label{eq:UNS}
{\partial\vv\over\partial t}
+\vv\cdot\nab\vv
+\frac 1 \rho \nab p
=\nu\nabla^2\vv,
\EN
where $p$ is pressure, is invariant under the transformation
\EQ
\xx\to\alpha\xx,\quad
\vv\to\alpha^h\vv,\quad
t\to \alpha^{1-h}t,\quad
\nu\to\alpha^{1+h}\nu.
\label{transfo}
\EN
In order that \Eq{dependence} can be satisfied, the exponents
$a$ and $b$ have to take certain values.
These exponents can be determined by requiring that $kt^a$ and $\nu t^b$
remain invariant under the scaling transformation \eq{transfo}, i.e.,
\EQ
kt^a\to (\alpha^{-1}k)[\alpha^{(1-h)}t]^{a}=kt^a,
\quad\text{so}\quad a={1\over1-h},
\EN
\EQ
\nu t^b\to(\alpha^{1+h}\nu)[\alpha^{(1-h)}t]^b=\nu t^b,
\quad\text{so}\quad b=-{1+h\over1-h}.
\EN
Translating this into a dependence on $q$ we have, using
$q=-(1+2h)$ and hence $h=-(q+1)/2$,
\EQ
a={2\over3+q},\quad
b=-{1-q\over3+q}.
\label{abcoefficientsq}
\EN
Note that $b=0$ for $q=1$, i.e., for initial energy spectra that
increase linearly with $k$.
In \Tab{tab:exponents} we have listed the scaling parameters for several
values of $q$.

The limit $t\to 0$ is problematic.
For $q>1$, i.e.\ $b>0$, both arguments of $g(x,y)$ vanish.
Assuming that $g(0,0)$ is finite, we can conclude that
for $t\to 0$ the dependence of $g(kt^a,\nu t^b)$
on $\nu$ and $k$ vanishes. This implies that
the zero point of $t$ corresponds to a time where the energy spectrum
would have been a pure power law,
\EQ\label{kq}
E(k,0,\nu)\sim k^q.
\EN
Such a spectrum is obviously singular and would have infinite energy.
We therefore refer to $t=0$ as a virtual zero point.
Near $t=0$ the spectrum can therefore not be self-similar.
On the other hand, if $g(x,y)$ is not necessarily finite in the limit $y\to0$,
the above conclusion cannot be made.
We return to this in \Sec{sec:SecondArgument}.
In the following we consider the case where time is sufficiently far
away from zero.

\begin{table} 
\begin{tabular}{cccccc}
\hline
~~~$q$~~~ & ~~~$h$~~~    & ~~~$a$~~~ & ~~~$b$~~~ & ~~~$2n$~~~ \\
\hline
1   & $-1.0$ & 1/2 & 0   & 1.00 \\
1.5 & $-1.25$& 4/9 & 1/9 & 1.11 \\
2   & $-1.5$ & 2/5 & 1/5 & 1.20 \\
3   & $-2.0$ & 1/3 & 1/3 & 1.33 \\
4   & $-2.5$ & 2/7 & 3/7 & 1.43 \\
\hline
\end{tabular}
\caption{Dependence of secondary scaling parameters on the slope $q$
of the initial spectrum.
The significance of $2n$ will be explained in \Sec{sec:SecondArgument}.}
\label{tab:exponents}
\end{table}
 
The validity of \Eqs{DecayingTurbulence}{dependence} has already been confirmed
using data from wind tunnel experiments \cite{CBC} where the viscosity is low
enough so that the second argument in $g(x,y)$, $y=\nu t^b$, can be neglected.
One goal of the present paper is to demonstrate, using direct simulations, that
\Eq{dependence} is also valid in the case where the second argument, $\nu t^b$,
cannot be neglected.  We do this by implementing in a numerical simulation a
time-dependent viscosity, $\nu=\nu(t)$, such that $\nu t^b=\const$ for a given
value of the initial power-law exponent $q$.

As a first test of the scaling
relationship we consider the decay of fields with initial power law
spectra; see \Eq{kq}.
We will then also test the scaling laws for
initial energy spectra that are not power laws (\Sec{sec:turbulent}).

\section{Comparison with simulations}

The Navier-Stokes equations for an isothermal
and weakly compressible fluid are solved in a box with periodic
boundary conditions.
We always adopt initial velocity fields such that their Mach number
is around 1\%, so compressibility effects can be neglected.
We employ the Pencil Code \cite{PencilCode} which
is a higher-order finite-difference code using the $2N$-RK3 scheme of Williamson
\cite{W80} for time stepping.  The low \Rey\  runs presented in this
paper were done on relatively coarse grids ($64^3$) 
while the high \Rey\ runs had a resolution of $256^3$.
For further details and recent turbulence simulations using
the Pencil Code see Ref.~\cite{DHYB03}.
We begin by studying the evolution of initial velocity fields
with power-law spectra.

\subsection{Initial power-law spectra}\label{sec:powerlaws}

The initial power-law spectra with arbitrary values of $q$ are
constructed by first generating in real space a random velocity field that is
$\delta$ correlated in space.
Such a velocity field corresponds to a $k^2$ energy spectrum.
In Fourier space, the velocity field $\vvv(\kk)$ is then multiplied
by a factor $k^{q/2-1}$.

\begin{figure}[h!]
  \centering
  \includegraphics[width=.5\textwidth]{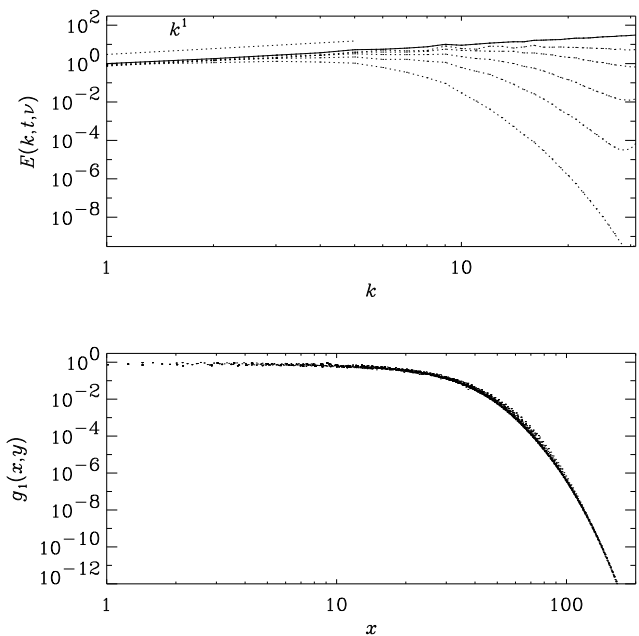}
  \caption{Decay of the initial energy spectrum $E(k,0,\nu)\sim k^q$
  for $q=1$ (upper panel) and the corresponding scaling function
  $g_q=E(k,t,\nu)/k^q$ versus $x=kt^{a}$ (lower panel).
  Note the collapse of the rescaled spectra for different times
  ($t=0$, 1, 2.5, 6, 12.5, and 29). For $q=1$ the parameter $a=1/2$.
  }
 \label{fig:k1}
\end{figure}

In the upper panel of \Fig{fig:k1} we show the results of a numerical
experiment where a power spectrum with $q=1$ decays under the action of
constant viscosity.
In this case we have $b=0$, so $y=\const$.
This means that the spectra collapse onto a single graph when
$E(k,t,\nu)$ is divided by $k^q$ ($=k$ in this case, because $q=1$)
and $k$ is multiplied by $t^a$ ($=t^{1/2}$ in this case).
This is indeed the case, see the lower panel of \Fig{fig:k1}.

\begin{figure}[h!]
 \centering
 \includegraphics[width=.5\textwidth]{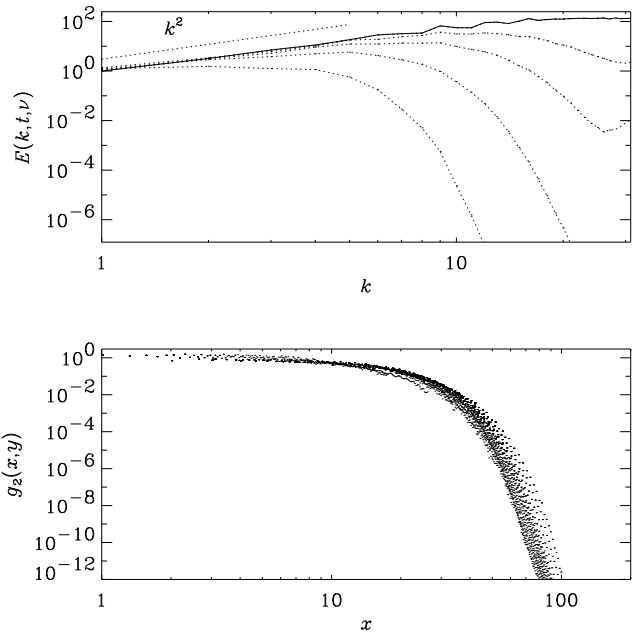}
 \caption{Same as \Fig{fig:k1}, but for $q=2$
 and constant viscosity, $\nu=10^{-3}$.
 The times shown in the upper panel are $t=0$, 3, 9, 26, and 77.
 Note that the curves for different times do not collapse onto a
 single graph.}\label{fig:k2qnu1} 
\end{figure}

For all other values of $q$, the second argument  $y=\nu t^b$  will not be
constant and must depend on $t$.
We therefore expect that the spectrum will not collapse onto a single graph.
This is shown in \Fig{fig:k2qnu1}, where we show the spectra at different
times (upper panel) and the attempt to collapse them onto a single graph
(lower panel).

\begin{figure}[h!]
 \centering
 \includegraphics[width=.5\textwidth]{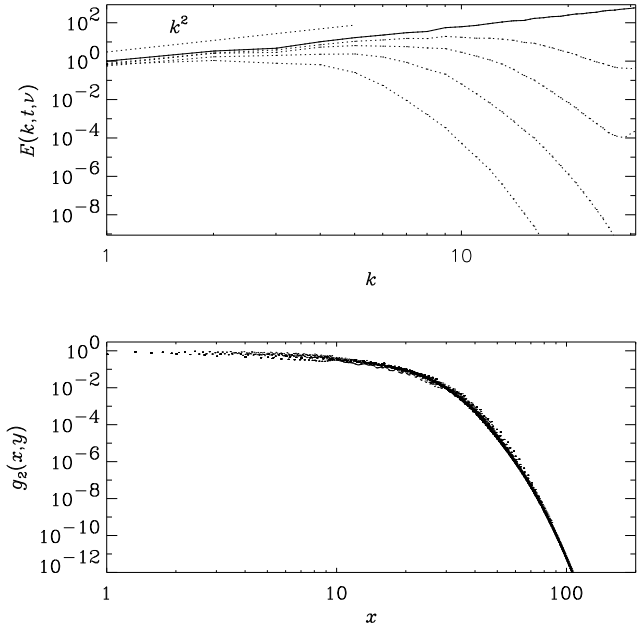}
 \caption{Same as \Fig{fig:k2qnu1}, but with $\nu=\nu(t)$ given
 by \Eq{eq:viscosity} with $\nu_{\rm ref}=3\times10^{-3}$ and
 $t_{\rm ref}=0.1$.
 Note that now the data points collapse reasonably well
 onto a single graph.
 In this figure the times are the same as in \Fig{fig:k2qnu1}.}
 \label{fig:k2qnu2} 
\end{figure}

Collapse of the spectra obtained at different times is in general
not possible unless one makes $\nu$ time dependent in such a way as
to keep $y=\nu(t)t^b$ constant in time.
In the following simulations the viscosity is therefore given by
\EQ\label{eq:viscosity}
\nu(t)=
\left\{ \begin{array}{ll}
\nu_{\rm ref} & \text{for $t\leq t_{\rm ref}$}\\
\nu_{\rm ref}  \left({t}/{t_{\rm ref}}\right)^{-b} & \text{otherwise},
\end{array}\right.
\EN
where $\nu_{\rm ref}=\nu(t_{\rm ref})$ is a constant reference viscosity.
At early times, $t<t_{\rm ref}$, the initial fields were allowed to
decay under the action of a constant viscosity $\nu_{\rm ref}$ so as
to avoid having to use an excessively large (or even infinite) viscosity.
The result is shown in \Fig{fig:k2qnu2} and the spectra for the different
times collapse reasonably well onto a single graph.

\subsection{Dependence of $g(x,y)$ on the first argument}
\label{sec:FirstArgument}

\begin{figure}[h!]
  \centering
 \includegraphics[width=.5\textwidth]{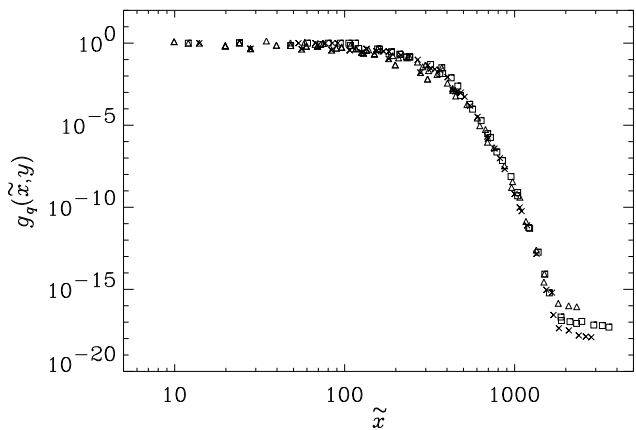}
 \caption{Three sets of $g(\tilde{x},y)$ curves from decay experiments with
 different values of $q$ but the same value of $y$ ($=2\times10^{-2})$.
 The abscissa has been rescaled according to \Eq{rescaling} to make
 the curves for $q=1$ (triangles), $q=2$ (squares), and $q=3$ (crosses)
 collapse onto a single graph.
 }\label{fig:gsamey}
\end{figure}

It turns out that for a fixed value of $y$ and 
different values of $q$ the scaling function $g(x,y)$ does not quite
collapse onto a single graph and that, therefore, the curves for
different values of $q$ are distinct.
We indicate this by a subscript $q$ and write $g_q(x,y)$.
However, empirically it turned out that to a good approximation
the $q$-dependence can be removed by rescaling $x$ by
a $q$ dependent factor, i.e.,
\EQ
x\to\tilde{x}=x(q+4)/5.
\label{rescaling}
\EN
Note that for $q=1$ we have $\tilde{x}=x$.
In \Fig{fig:gsamey} we show $g_q(\tilde{x},y)$ versus $\tilde{x}$
for fixed value of $y$ and three different values of $q$ (=1, 2, and 3).
Note that the collapse of the three curves is reasonably good.

\subsection{Modified time dependence of viscosity}

\begin{figure}[h]
\centering
\includegraphics[width=.5\textwidth]{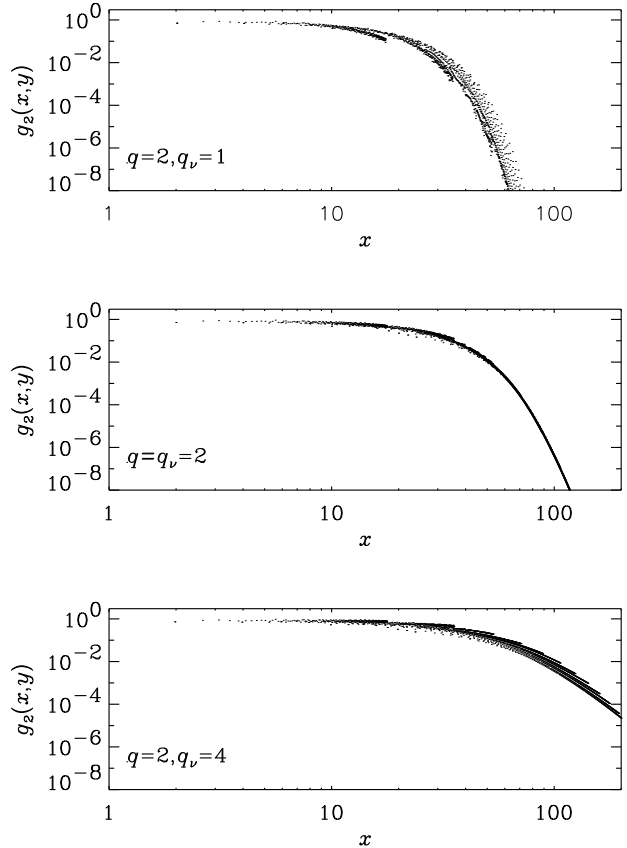}
 \caption{Scatter plots for $q=2$ and different
 values of $q_\nu$.
 Note that the best collapse is achieved for $q_\nu=q$.
 }\label{fig:multiq2}
\end{figure}

In order to verify the anticipated scaling behavior further,
we determine effective values of $q$ and check whether these
values are consistent with each other.
We begin by defining an effective value $q_\nu$ that determines
the time dependence of $\nu(t)$.
Thus, $\nu(t)$ is proportional to $t^{b_\nu}$ where
$b_\nu=-(1-q_\nu)/(3+q_\nu)$, which is analogous to \Eq{abcoefficientsq}.
The result is shown in \Fig{fig:multiq2} and we see that the best collapse
is indeed achieved when $q=q_\nu$.
We have checked that this agreement holds also for different values of $q$.

\subsection{Dependence of $g_q(x,y)$ on the second argument}
\label{sec:SecondArgument}

Next we consider the temporal decay law of the kinetic energy.
This allows us to constrain the dependence of $g(x,y)$ on $y$.
As usual, the kinetic energy (per unit mass and unit volume)
can be found by integration,
\EQ
E_{\mathrm{kin}}(t,\nu) = \int_0^\infty E(k,t,\nu)\,\dd k
\quad(t>0).
\label{EnergyIntegral}
\EN
For a given value of $y$, where $y$ may still be a function of $t$,
\Eq{EnergyIntegral} can be rewritten as an
integral over the first argument of the scaling function, $x=kt^a$.
This gives
\EQ
E_{\mathrm{kin}}(t,y)=t^{-a(1+q)}\int_0^\infty x^q g_q(x,y)\,\dd x,
\EN
where $E_{\mathrm{kin}}$ still depends on $y=y(t)$.
Here we have ignored the fact that in order for $g_q(x,y)$ to be
independent of $q$ we should rescale the $x$ coordinate by a factor
$(q+4)/5$; see \Eq{rescaling}.
However, this only corresponds to an overall rescaling of the kinetic
energy by a factor $[(q+4)/5]^{-(q+1)}$ and is therefore unimportant.

It is convenient to isolate the main $t$ dependence,
$E_{\mathrm{kin}}\sim t^{-2n}$, where $2n=a(1+q)$, and
\EQ
n={1+q\over3+q}.
\EN
We can therefore write
\EQ\label{decay_law}
E_{\mathrm{kin}}(t)\sim  t^{-2n}\,\tilde{g}_q(\nu t^b),
\EN 
where $\tilde{g}=\tilde{g}(y)$ is a function that only depends
on one argument and is obtained by integrating out the $x$-dependence
of $g(x,y)$, i.e.\
\EQ
\tilde{g}_q(y)=\int_0^\infty x^q g_q(x,y)\,\dd x.
\EN
The resulting values of $2n$ are given in \Tab{tab:exponents}
for different values of $q$.
Note that the basic $t^{-2n}$ decay law has also been obtained in
Ref.~\cite{DJO}, but there it was assumed that $\nu$ is negligibly small
and that for large values of $k$ one has a Kolmogorov spectrum.
The basic relation between $2n$ and $h$ or $q$ can also be obtained
by assuming that the rate of dissipation is proportional to
$v_\ell^3/\ell$, where $\ell$ is the integral scale \cite{Frisch}.

\begin{figure}[h]
\centering
\includegraphics[width=.5\textwidth]{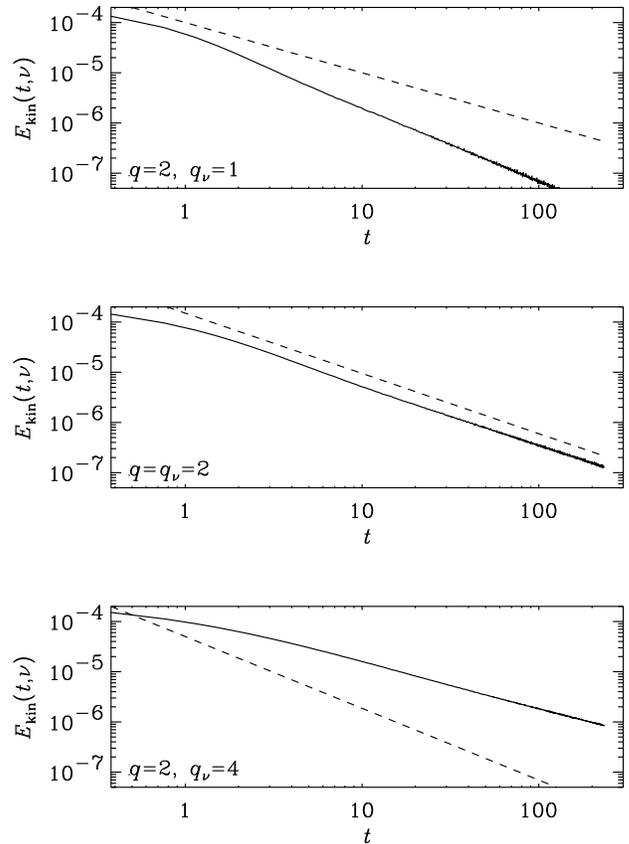}
 \caption{Energy decay law for $q=2$ and different
 values of $q_\nu$ (solid lines).
 The different slopes are $2n=1.5$, 1.2, and 0.9.
 The dashed lines indicate the slope expected if
 the decay law was governed by the value of $q_\nu$
 (slopes 1, 1.2, and 1.43; see \Tab{tab:exponents}).
 Again, the best collapse is achieved for $q_\nu=q$,
 corresponding to $2n=1.2$.
 }\label{fig:decayq2}
\end{figure}

In \Fig{fig:decayq2} we check that the basic decay law is indeed mostly
governed by the value of $q$ and not by the value of $q_\nu$.
For $q_\nu=q=2$ the decay has the expected slope $2n=1.2$ (middle panel).
For $q_\nu=1$, the viscosity is constant and the decay is accelerated,
while for $q_\nu=4$, the viscosity decreases faster than is necessary
for keeping $y$ constant, and the decay of $E_{\mathrm{kin}}$ is now
slower than what is expected based on the value of $q$.
These results confirm that the best agreement is achieved for $q_\nu=q$.
We have checked that the same is true for $q_\nu=q=3$, for example.

\begin{figure}[h]
\centering
\includegraphics[width=.5\textwidth]{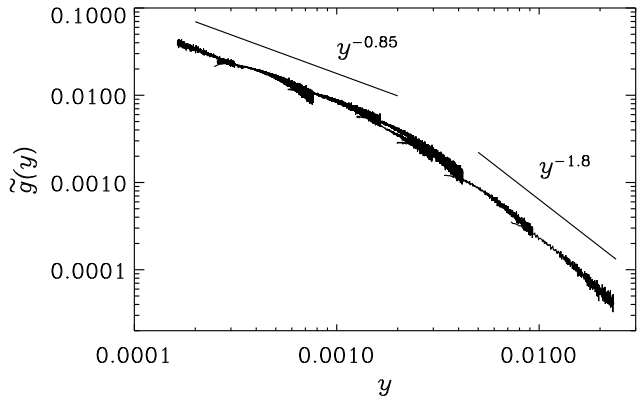}
 \caption{
Representation of $\tilde{g}_q(y)$ obtained by patching together the
decay laws from different runs with different values of $q$.
At early times the decay is not yet self-similar, so the data for
these times have been ignored in the plot (the remains of the
initial transients can still be seen as little hooks in the beginning
of each piece).
Each piece of the decay curve has been shifted along the ordinate
to connect the different pieces with each other.
The normalization of $\tilde{g}$ is therefore arbitrary.
}\label{fig:pgtilde}
\end{figure}

We now turn to the $y$ dependence of $\tilde{g}_q(y)$,
which can be determined by plotting
$t^{2n}E_{\mathrm{kin}}$ versus $y$; see \Fig{fig:pgtilde}.
Within plotting accuracy the results seem to be
independent of the value of $q$.
We can therefore drop in the following the subscript $q$ on $\tilde{g}_q(y)$.

The results confirm that for small values of $\nu$ 
the time dependence of the decay law of kinetic energy is weaker: 
$\tilde{g}\sim y^{-0.85}$ for
$y<0.003$ compared to $\tilde{g}\sim y^{-1.8}$ for larger values of $y$.
However, there is as yet no evidence that $\tilde{g}$ becomes completely
independent of $y$ when $y\to0$.

The above results imply that the energy decay law is attenuated
by a small correction factor for $q\neq1$.
Consider, as an example, the case $q=2$.
The basic decay law is $E_{\rm kin}\sim t^{-1.2}$, see \Eq{decay_law}.
For $q=2$ we have $b=1/5$, so for small values of $\nu$
(assuming $y<0.003$) the exponent has to be corrected by $-0.85/5=-0.17$,
so that the correct decay law is $E_{\rm kin}\sim t^{-1.37}$.
This is indeed confirmed by direct inspection of the data.

The fact that $\tilde{g}$ does not seem to go to a finite value in
the limit $y\to0$ is surprising, because it implies that there is a
viscosity correction to the basic $t^{-2n}$ decay law even in the limit
of vanishing viscosity.
Although the data do not necessarily allow such an extrapolation, we
are not aware of any evidence against a finite viscosity correction in
the large Reynolds number limit.

\subsection{Turbulent initial conditions}
\label{sec:turbulent}

The results shown in the previous sections demonstrate that the scaling
law \eq{DecayingTurbulence} successfully describes the decay of kinetic
energy in the special case of a flow field with an initial $k^q$ spectrum.
A more realistic initial condition is a turbulent velocity field which has
an energy spectrum that is decreasing with increasing $k$.
Nevertheless, there is always a subinertial range where values of $q$
between 1 and 4 are not uncommon.

\begin{figure}[t!]
  \centering
  \includegraphics[width=.5\textwidth]{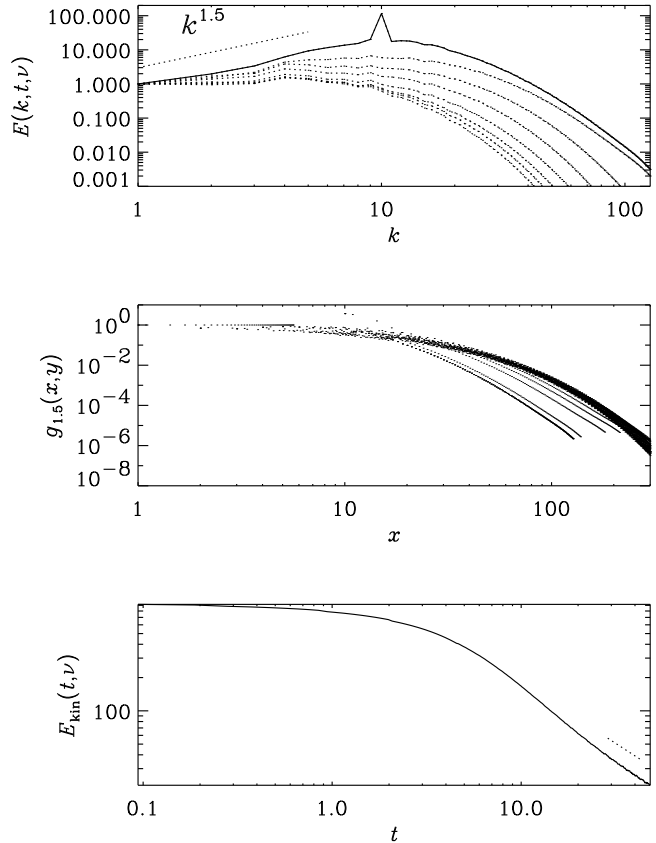}
  \caption{
 Upper panel: initial energy spectrum (solid line) together with subsequent
 energy spectra (dotted lines) obtained from driven turbulence simulation
 forced at wave number $k_{\rm f}=10$.
 The spectral slope for small wave numbers corresponds to $q=1.5$.
 Middle panel: attempt to collapse the spectra on a single graph
 which fails at early times.
 Lower panel: decay of kinetic energy corresponding to a slope
 $2n=10/9\approx1.11$, confirming the $q=1.5$ scaling.
 }\label{forcedk10t0_0}
\end{figure}

In the following we consider the decay of flow fields
that are initially statistically stationary. These initial fields are produced
by applying a random force within a band of wave numbers around $k_{\rm f}$
until the work done by the forcing is balanced by dissipation.
Relatively high resolution ($256^3$) and large values of $k_{\rm f}$
are needed in order to get a well-defined subinertial range.

As explained in \Sec{sec:decaying}, the scaling law \[\Eq{dependence}\]
is a direct consequence of the scaling properties of the unforced Navier-Stokes
equations,
\Eq{eq:UNS}, and will not be valid for a flow driven by a general forcing
function.
The statistically stationary state considered here will not necessarily
be compatible with the Navier-Stokes equations.
In the following we assume that $t=0$ is the time when the forcing is
stopped, but we must expect there to be some readjustment phase before
self-similar scaling is possible.

When viscosity can be considered negligible or when it is made
time dependent according to \Eq{eq:viscosity}, the parameter $q$
can, in principle, be determined by two independent methods.  
It can be found
by determining the spectral slope in the subinertial range
or by fitting the energy decay to \Eq{decay_law}, as done in Ref.~\cite{DJO}.
In addition, of course, $q$ could be found completely empirically by
trying different values until the collapse is best.
Unfortunately the first approach is difficult since one has to have large
scale separation between the size of the box and the integral (or forcing)
scale in order to be able to resolve the infrared region.
This requires very large resolution.
In addition, the infrared limit obtained from simulations is not very
accurate for small $k$, because only a few modes contribute to the
shell-integrated spectrum.

\begin{figure}[t!]
  \centering
  \includegraphics[width=.5\textwidth]{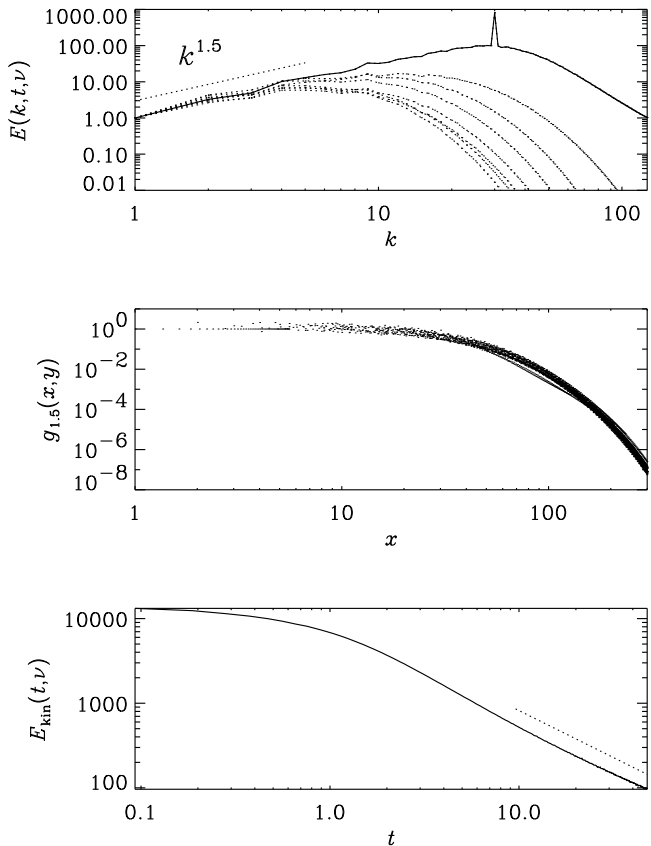}
  \caption{
 Same as \Fig{forcedk10t0_0}, but for $k_{\rm f}=30$. Note that the
 collapse is much better -- even at earlier times -- than for the case
 with $k_{\rm f}=10$ (middle panel).
 }\label{forcedk30t0_0}
\end{figure}

In \Fig{forcedk10t0_0} we show the result for a turbulence simulation
that was driven at $k_{\rm f}=10$ and the forcing was turned off at $t=0$.
Note that the collapse is relatively poor at early times.
The collapse improves significantly when the turbulence is
driven at $k_{\rm f}=30$; see \Fig{forcedk30t0_0}.

The reason for the collapse being much better in the case of larger
$k_{\rm f}$ is probably related to the facts that the local turnover time
$\tau_k\sim(u_{\rm rms}\ell)^{-1}$ is shorter.
Thus, self-similarity can probably commence much earlier.

Finally, we note that in our simulations the subinertial range slope is
$q=1.5$ both for large and small
values of $k_{\rm f}$.  We are not aware of a theoretical explanation for this
slope, but it is probably related to finite size effects.  By contrast, in an
infinite domain the slope is expected to be $q=4$ (or $=2$), which could be
motivated if the Loitsyansky (or Saffman) integral were independent of time.
In that case one would have $2n=10/7$  (or $2n=6/5$).

\subsection{Conclusion}

The results presented above have shown that decaying hydrodynamic
turbulence can be characterized by a two-parametric scaling function and
that this function may well be universal and independent of the initial
slope $q$ of the spectrum in the infrared limit, i.e.\ in the subinertial
range.
Although the basic scaling behavior has already been confirmed earlier
\cite{DJO}, using data from wind tunnel turbulence \cite{CBC}, it was
not possible to determine the infrared scaling properties of the energy
spectrum.
Indeed, for the smallest wave numbers available from the wind tunnel
data the spectrum was still a decreasing function of wave number $k$.
This is because wind tunnel measurements only allow access to
one-dimensional spectra which are always monotonically decaying.
This property follows from the fact that for isotropic turbulence the
three-dimensional spectrum $E(k)$ is related to the one-dimensional
spectrum $E_{\rm 1D}(k)$ via $E(k)=-k\dd E_{\rm 1D}/\dd k$, and since
$E(k)>0$ it follows that the one-dimensional spectrum can never increase
with $k$; see Eq.~(7) of Ref.~\cite{DHYB03}.
This is also true for the longitudinal and transversal power spectra
separately; see, for example, Fig.~6.11 of Ref.~\cite{Pope00}.
Whether or not the proper subinertial range of the three-dimensional
spectrum can be determined from wind tunnel experiments is unclear.
It is therefore important that simulations can now demonstrate explicitly
that the slope of the subinertial range spectrum is linked to the scaling
law derived in Ref.~\cite{DJO}.

There are obvious extensions of this work to the case of decaying
magnetohydrodynamics turbulence.
Similar scaling properties also apply to the magnetic case \cite{O97,KP},
but the detailed functional dependence of the corresponding two-parametric
scaling function has not yet been fully determined, although partial
results do already exist.
In particular, for the case of helical initial fields the combined dependence
on wavenumber and time has been studied in Ref.~\cite{CHB01}, and resistive
corrections to the decay law have been investigated in Ref.~\cite{CHB02}.
This work generalizes earlier findings that in the helical case the magnetic
energy can decay as slowly as $\sim t^{-1/2}$ \cite{BM}, while in the
nonhelical case the decay is generally faster and similar to the hydrodynamic
case \cite{MacLow}.

A general difficulty with the self-similarity  approach is the uncertainty
regarding the zero point of $t$.
There is apparently no unique way of determining this time {\it a priori}.
An {\it a priori} choice of the zero point of $t$ is however necessary if
$\nu$ is allowed to be a function of time.
Although the uncertainty regarding the zero point of $t$ becomes less
influential at later times, it is not normally possible to revise the
zero point of $t$ afterwards, unless one is prepared to run an entirely
new simulation.

Once the initial startup phase is over and the decay has become
self similar, one is however able to determine in full detail the exact
form of the two-parametric scaling law.
Our current work can only be preliminary, because it remains to be
checked how general and perhaps even universal the $g(x,y)$ function
really is.
If its generality is established, it could become a powerful analysis
tool for making predictions about the decay of kinetic energy.
This applies in particular to the function $\tilde{g}(y)$, which plays
the role of a viscosity dependent correction function for the decay law
of the kinetic energy.

\acknowledgments
We acknowledge the use of the parallel computers in Trondheim, Odense, and
Bergen.

\end{document}